%% file: main.tex
\documentclass{article}

\newif\ifanon

\ifanon
  \usepackage{neurips_2026}            
\else
  \usepackage[preprint]{neurips_2026}  
\fi

\usepackage[utf8]{inputenc}   
\usepackage[T1]{fontenc}      
\usepackage{hyperref}         
\usepackage{url}              
\usepackage{booktabs}         
\usepackage{amsmath}          
\usepackage{amsfonts}         
\usepackage{amssymb}          
\usepackage{nicefrac}         
\usepackage{microtype}        
\usepackage{xcolor}           
\usepackage{graphicx}         
\usepackage{bbm}

\newcommand{\hforg}{https://huggingface.co/the-mitre-corporation}

\ifanon
  \newcommand{\available}{Our fine-tuned models and dataset will be released publicly upon publication; the repository is withheld for anonymity.}
\else
  \newcommand{\available}{Our fine-tuned models and dataset are publicly available at \url{\hforg}.}
\fi

\title{BUSTA: Teaching Small Language Models to Reason About Binary Execution}

\author{%
  \textbf{Ritika Poddar}$^{*}$, \textbf{Aidan Fennelly}$^{\dagger}$, \textbf{Deirdre Murphy}, \\[3pt]
  \textbf{Eric Wolos}$^{\dagger}$, \textbf{Corre L.\ Steele}, and \textbf{Michael Doyle}$^{*}$ \\[5pt]
  The MITRE Corporation \\[3pt]
  $^{*}$Corresponding authors: \texttt{rpoddar@mitre.org}, \texttt{mdoyle@mitre.org} \\
  {\small $^{\dagger}$Work done while at The MITRE Corporation; for contact, see the corresponding authors.}
}

\begin{document}

\maketitle

\input{sections/01_abstract}

\input{sections/02_introduction}
\input{sections/04_method}
\input{sections/05_experiments}
\input{sections/06_conclusion}

\input{sections/08_limitations_future_works}
\bibliographystyle{plainnat}
{\small \bibliography{references}}

\paragraph{Availability.}
\ifanon
\else
We release our fine-tuned models and training dataset under the Apache-2.0 license via the \url{\hforg} organization on HuggingFace.
\fi

\ifanon\else
\begin{ack}
We acknowledge Gabby Raymond, Kyle Skey, Peter Lucia, and Tim Welsh for their guidance during the experimental design and editing.

This work was supported by the MITRE Independent Research and Development program.  © 2026 The MITRE Corporation.  All rights reserved.  Approved for public release.  Distribution unlimited.  Case 26-1796.
\end{ack}

\paragraph{Author Contributions.}
A.F.\ and R.P.\ built the training and evaluation pipeline. D.M.\ constructed
the initial dataset, and A.F.\ led the final data processing and labeling. C.S.\ provided reverse-engineering domain expertise that guided the
dataset design and reward function, and contributed to writing. E.W.\ initiated the training experiments.
R.P.\ and M.D.\ led the analysis and writing.
\fi

\appendix
\input{sections/07_appendix}


\end{document}

%% file: sections/01_abstract.tex
\begin{abstract}

  Large Language Models (LLM) are increasingly applied to Reverse Engineering (RE) workflows. RE, however, differs from common LLM coding tasks as it requires models to reason about low-level program behavior. Furthermore, the deployment of tooling in resource-constrained sandboxed RE environments limits the ability to use frontier-scale solutions. We examine whether a targeted LLM training approach can transfer distinct RE skills to open models via inverse basic-block execution: given an x86-64 basic block and a desired output register state, the model predicts the input register values that produce it. Fine-tuning \texttt{Qwen3-1.7B} and \texttt{Nemotron-3-Nano} with Group Relative Policy Optimization (GRPO) raises knowable-register prediction accuracy from 73.2\% to 85.3\% and from 63.3\% to 95.2\%, respectively. Frontier models still score highest (up to 99.2\%), but our fine-tuned \texttt{Nemotron-3-Nano} comes within four points at a fraction of the parameter count. We release both fine-tuned models and the training dataset.
\end{abstract}

%% file: sections/02_introduction.tex
\section{Introduction}
\label{sec:intro}

Reverse engineering (RE) is integral to vulnerability research and malware analysis~\citep{manuel2024enhancing, sok_llm_re2025, rex86_2025}, and while momentum for LLM-based RE research is growing, zero- and few-shot prompting approaches are far more represented than fine-tuning~\citep{sok_llm_re2025}. RE tasks are not always readily generalizable from other common software tasks. RE has its own syntax; useful models must read assembly, registers, and traces, as well as reason over low-level program state~\citep{rex86_2025}. We therefore study whether targeted training can teach models useful RE skills that general prompting does not.

We present a novel application determining what program inputs or initial machine state lead to a desired execution state, a task that arises frequently in vulnerability research. Symbolic execution engines such as \texttt{angr} can sometimes solve this problem by exploring execution paths and solving the resulting constraints \citep{shoshitaishvili2016state,stephens2016driller,shoshitaishvili2015firmalice}, but are often limited by path explosion and expensive constraint solving \citep{cadar2013symbolic,baldoni2018survey}. Recent work pairs LLMs with symbolic and concolic execution to guide search, construct harnesses, or assist constraint solving \citep{sailor,autobug,cottontail-sp26,nong2026neurosymbolicproofofvulnerabilitygenerationopenweight,eslamimehr2026hybridconcolictestinglarge,SymTEE,kleecopilot}, but these approaches primarily operate over source code or source-level intermediate representations. We instead investigate whether LLMs can reason directly over disassembled binary code when source is not available. As a first step, we study a simplified basic-block problem: given a basic block and desired output register values, can an LLM predict input register values that produce them? 

Our secondary focus is whether a small fine-tuned model can rival out-of-the-box frontier models. Reverse-engineering workflows frequently analyze sensitive or malicious binaries in isolated environments where third-party frontier APIs cannot be used; open-weight models can instead run locally~\citep{rex86_2025}, and among these, smaller models fit tighter compute budgets~\citep{samsi2023watts}.

We fine-tune two models with our approach, BUSTA (Binary UnderStanding Task Assistant): \texttt{BUSTA-BB-Qwen3-1.7B} and \texttt{BUSTA-BB-Nemotron-3-Nano}, using Group Relative Policy Optimization (GRPO)~\citep{deepseekmath2024} on inverse basic-block runtime prediction, and demonstrate significant gains in knowable-register accuracy over the base models, with \texttt{Qwen3-1.7B} increasing from 73.2\% to 85.3\% and \texttt{Nemotron-3-Nano} from 63.3\% to 95.2\%.\footnote{\available} We comprehensively benchmark across open and frontier models on a diverse dataset, with significant effort toward equitable and reliable comparison across differing model generation styles.

%% file: sections/04_method.tex
\section{Method}
\label{sec:method}


\subsection{Data Curation}

We curate our dataset from Assemblage~\citep{assemblage2024}, a corpus of x86-64 Executable and Linkable Format (ELF) and Windows Portable Executable (PE) binaries. From 100 Windows PE binaries, we use a Ghidra~\citep{ghidra} script to generate disassembly and segment it with Ghidra's \texttt{SimpleBlockModel} into 287,137 basic blocks---distinct fixed sequences of machine instructions bounded at control-flow-changing instructions such as calls, branches, or jumps.

We then emulate each block using Unicorn~\citep{unicorn} with randomized initial register values, except for \texttt{RIP}, \texttt{RSP}, and \texttt{RBP}, which we fix as the pointers for the execution. We log the input and output register states. Each resulting training and evaluation data sample contains a single execution of a basic block, including the block address, the static disassembly, and its input and output register states.

As part of preprocessing, we also label registers in each basic block for which multiple input values could produce the same output state. The value of these labeled input registers cannot be determined from examining only the basic block and an output state. To generate this label, we examine whether two or more different initial values for a register could produce the same final register state. We use an \texttt{angr}~\citep{shoshitaishvili2016state} script to initialize all registers and memory as symbolic, execute the basic block twice, constrain the outputs of each execution to be equal, and sequentially constrain each pair of matching input registers to be different. If this constraint is solvable, then two different initial values for that register could produce the same final register state, in which case that register is marked as ``unknowable.'' All other registers are marked as ``knowable.'' \texttt{RIP}, \texttt{EIP}, \texttt{IP}, and \texttt{RFLAGS} are always treated as unknowable. 

After applying several validation filters and structuring the basic block sample objects for clear register labeling (Appendix~\ref{app:data}), we retain 282,741 valid samples, split 80/10/10 into training, validation, and test sets. The sets are mutually exclusive with respect to the functions from which each basic block is derived. We further downsample the training pool via stratified sampling to preserve diversity while shrinking it (Appendix~\ref{app:down}).


\subsection{Training Method}

We perform reinforcement learning (RL) post-training using GRPO, an on-policy algorithm suitable for optimizing models against sparse task-level reward signals. Training is implemented with the open-source NeMo RL post-training library~\citep{nemorl}, using the Megatron~\citep{megatronlm2019} backend. To guide optimization, we define a custom reward function composed of multiple weighted objectives.

To constrain the structured generation, training prompts include Pydantic~\citep{pydantic} prompt schemas that explicitly outline the input and output expected for model predictions.


\subsection{Reward Function}

The reward function scores the model's predicted register state against ground truth. The reward is a weighted combination of a register-value score $S_{\text{reg}}$ and an unknowable-classification score $S_{\text{unk}}$:
\begin{equation}
  S_{\text{final}} = 0.7\,S_{\text{reg}} + 0.3\,S_{\text{unk}}.
\end{equation}

Both components are computed by averaging per-register rewards and applying a $2^{x}-1$ transform. Since accurately recovering knowable register values is the primary objective, $S_{\text{reg}}$ receives greater weight. Correct and incorrect values for knowable registers receive rewards of $+1$ and $-1$, respectively, while correct values for unknowable registers receive a $+1.1$ reward without penalty for incorrect values. The auxiliary $S_{\text{unk}}$ term rewards or penalizes each predicted register by $+1$ or $-1$ according to whether its knowable/unknowable label is correct. The model is expected to predict all registers present in the output register state, and omitted ground-truth registers incur a $-0.5$ penalty each. Invalid JSON or schema-noncompliant responses receive a fixed $-0.5$ penalty as the $S_{\text{final}}$ score. We select the weights and penalties through hyperparameter experimentation; the full per-register formulation appears in Appendix~\ref{app:reward}.


\subsection{Evaluation}

For evaluation, we downsample the held-out test split by capping each stratum (defined in Appendix~\ref{app:down}) at 100 samples, yielding 635 samples. Our evaluation framework benchmarks four primary metrics: valid response rate, register coverage, knowable register accuracy, and unknowable classification accuracy. We include unknowable value accuracy as a supplemental metric. 
Beyond accuracy, a central motivation of this work is the compute footprint of models that must run locally in constrained RE environments, where frontier APIs are often unavailable. We examine the mean total token usage per response for each model, including both input and output tokens, as a deployment-agnostic proxy for inference cost.

We evaluate our best performing \texttt{BUSTA-BB-Qwen3-1.7B} and \texttt{BUSTA-BB-Nemotron-3-Nano} models against the respective base models \texttt{Qwen3-1.7B} \citep{qwen3_2025} and \texttt{Nemotron-3-Nano} \citep{nemotron3nano2025}, and frontier models including \texttt{GPT-5} \citep{openai_gpt5_2025}, \texttt{GPT-5.5} \citep{openai_gpt55_2026}, \texttt{Claude-Opus-4.8} \citep{anthropic_opus48_2026}, \texttt{Claude-Sonnet-5} \citep{anthropic_sonnet5_2026}, \texttt{Gemma-4-E2B}, and \texttt{Gemma-4-31B} \citep{gemma4_2026}. To ensure fair comparison, we implement a comprehensive evaluation protocol. We tune temperature and top-p per model. We also utilize the vLLM~\citep{vllm2023} reasoning parser to separate reasoning from structured output, and conduct response post-processing. Final results are aggregated from multiple seed runs. Full protocol details are in Appendix~\ref{app:eval}.

%% file: sections/05_experiments.tex
\section{Experiments}
\label{sec:experiments}

We begin our experiments by fine-tuning \texttt{Qwen3-1.7B} on the inverse register inference task using GRPO with a 32K-token context window and a batch size of 512. We select this relatively small model to stay within the compute constraints of our target deployment environment and enable rapid iteration.

We perform a hyperparameter search by varying the invalid-response penalty, the per-stratum sample cap, and the reward-metric weights (Appendix~\ref{app:train}). The best \texttt{Qwen3-1.7B} configuration doubles the base per-stratum cap to 4k samples, yielding 22,223 training and 2,745 validation samples; it uses an invalid-response penalty of $-0.5$ and completes a full epoch (347 steps).

Following the success of this configuration, we apply the same training, sampling, and reward settings to \texttt{Nemotron-3-Nano-30B-A3B-BF16}. We fine-tune Nemotron on the inverse register inference task using GRPO and Low-Rank Adaptation (LoRA)~\citep{lora2021}, with rank-128 adapters and a scaling factor of $\alpha=512$, on a four-node non-collocated architecture with separate vLLM rollout and Megatron optimization workers. Stabilizing this setup requires importance-sampling correction, sequence masking, a Kullback--Leibler (KL) divergence penalty, and learning-rate adjustments to complete the full 347-step run; we detail these in Appendix~\ref{app:train}.

\begin{table}[t]
\centering
\small
\setlength{\tabcolsep}{4pt}
\renewcommand{\arraystretch}{1.1}
\caption{Basic-block input register prediction results. \textit{Valid} is the fraction of responses that
parsed. \textit{Cov.}\ is register coverage. The accuracy columns are knowable-register,
unknowable-classification, and unknowable-value accuracy. \textit{Avg. Tot. Tokens} is the average tokens per response. $\Delta$ is the change in knowable-register
accuracy relative to the corresponding base model.}
\label{tab:consolidated-results}
\begin{tabular}{l r r r r r r r}
\toprule
& & & \multicolumn{4}{c}{Accuracy (\%)} & \\
\cmidrule(lr){4-7}
Model & Valid (\%) & Cov.\ (\%) & Known & $\Delta$ & Unk.\ cls. & Unk.\ val. & Avg. Tot. Tokens\\
\midrule
\multicolumn{8}{l}{\textit{Open-weight models}} \\
\texttt{qwen3-1.7b}             & 97.2  & 77.8  & 73.2 & --      & 59.5 & 6.3 & 14716 \\
\quad \texttt{+ busta}          & 94.1  & 89.4  & 85.3 & $+12.1$ & 72.8 & 10.2 & 7668\\
\texttt{nemotron-3-nano}        & 72.0  & 68.3  & 63.3 & --      & 58.4 & 9.3  & 27968\\
\quad \texttt{+ busta}          & 99.8  & 99.5  & 95.2 & $+31.9$ & 91.2 & 11.8 & 15794\\
\texttt{gemma-4-E2B}            & 100.0 & 86.7  & 63.7 & --      & 73.1 & 6.9  & 5671\\
\texttt{gemma-4-31B} & 99.9  & 99.9  & 98.1 & --      & 91.2 & 34.6 & 12654\\
\midrule
\multicolumn{8}{l}{\textit{Frontier API models}} \\
\texttt{claude-sonnet-5} (high) & 100.0 & 100.0 & 99.0 & -- & 94.3 & 38.4 & 14031\\
\texttt{opus-4.8}               & 100.0 & 99.9  & 98.1 & -- & 94.4 & 38.8 & 9062\\
\texttt{gpt-5}                  & 100.0 & 99.8  & 98.3 & -- & 93.4 & 39.0 & 12307\\
\texttt{gpt-5.5} (medium)       & 100.0 & 100.0 & 99.2 & -- & 94.8 & 39.3 & 7232\\
\bottomrule
\end{tabular}
\end{table}

\subsection{Results}
Fine-tuning substantially improves both open-weight base models (Table~\ref{tab:consolidated-results}); knowable-register accuracy rose 12.1 points to 85.3\% for \texttt{BUSTA-BB-Qwen3-1.7B} and 31.9 points to 95.2\% for \texttt{BUSTA-BB-Nemotron-3-Nano}, with comparable gains in unknowable classification. The average total tokens also reduced substantially, with \texttt{BUSTA-BB-Qwen3-1.7B} using nearly half the tokens of the base model. The valid response rate also increased significantly for \texttt{BUSTA-BB-Nemotron-3-Nano} to 99.8\%, however it slightly decreased for \texttt{BUSTA-BB-Qwen3-1.7B} to 94.1\%. Among open-weight comparisons, \texttt{Gemma-4-31B} is strong at 98.1\%, while the small \texttt{Gemma-4-E2B} reaches only 63.7\%, a plausible future fine-tuning target. While frontier models remain strongest overall, led by \texttt{GPT-5.5} at 99.2\%, and hold a clear edge on the harder bonus unknowable-value recovery, the fine-tuned BUSTA models close most of the primary-task gap at far smaller scale.

%% file: sections/06_conclusion.tex
\section{Conclusion}
\label{sec:conclusion}

Fine-tuning small open-weight models with GRPO on inverse basic-block execution substantially improves knowable-register accuracy, evidence that targeted post-training can improve performance for tasks requiring low-level binary reasoning. Though frontier models remain at the top for knowable-register accuracy, our experimentation closes much of the gap to out-of-the-box frontier models at a significantly smaller scale, suggesting that targeted post-training can be a practical option for compute-constrained reverse-engineering environments.

%% file: sections/08_limitations_future_works.tex
\section{Limitations and Future Work}
\label{sec:limitations}

Our experiments have several limitations that suggest directions for future work. First, the dataset is limited to 100 binaries from Assemblage; future work should evaluate on a larger and more diverse set of unseen binaries. Reward design is another potential area for improvement; additional iterations could better handle edge cases and more clearly distinguish knowable-register accuracy from broader register-classification performance.

Future work could also revisit the training task formulation. In early experiments, we trained on both the forward and inverse runtime prediction tasks, hypothesizing that forward prediction might improve code understanding and thereby improve inverse prediction. We ultimately removed the forward task to focus solely on inverse prediction, but its role remains worth further analysis. Finally, this work could be extended beyond individual basic blocks to sequences of basic blocks and, ultimately, whole-program target-state reachability. Evaluating in the actual target deployment environment would further assess practical utility and robustness.

%% file: sections/07_appendix.tex
\section{Data Curation Details}
\label{app:data}

During initial data processing, each basic block is emulated with Unicorn three times. We then apply several validation filters to identify invalid samples:
\begin{enumerate}
  \item A block is removed when Unicorn emulation fails or when parsing produces inconsistent traces.
  \item A sample is marked invalid if instruction addresses in the register trace do not exactly match the instruction addresses in the assembly, or if memory-trace addresses are not a subset of the block's instruction addresses.
  \item A sample is also flagged when memory-trace presence is inconsistent across runs, since a valid block should either always or never produce memory traces across its three executions.
\end{enumerate}
Each data sample is then reduced to the first of the three runs so that each sample contains the results of a single run. We also omit per-instruction memory and register traces from the data samples prior to training and evaluation as they provide additional context about the register state changes and are generally unavailable in realistic RE workflows.

To ensure code understanding by the model, we enhance the trace object structure so that the input and output registers are explicitly labeled, with clearly distinguishable register key names and values. This leaves us with 282,741 valid samples, split into 80\% training (225,993), 10\% validation (28,374), and 10\% test (28,374) sets. 

\section{Downsampling}
\label{app:down}

We downsample using stratified sampling to preserve diversity while reducing sample size. We partition samples into 8 strata defined as the cross-product of a binary memory-trace indicator and a four-way instruction-count bin (1--3, 4--7, 8--15, and 16+ instructions); each basic block is assigned to exactly one stratum by its (has-memory-trace, instruction-count-bin) pair. Within each stratum, downsampling first greedily retains samples that introduce previously unseen mnemonics---preserving rare mnemonics---before randomly filling the remaining slots up to a per-stratum cap; strata smaller than the cap are kept whole. The resulting set is thus enriched for memory traces, high instruction counts, and rare mnemonics. Although we retain the samples containing memory-access instructions, we remove the per-instruction memory traces from each sample prior to training and evaluation to avoid leaking register-state information to the model.

The same capping mechanism is applied at two independent stages. To build the held-out evaluation set, we cap the test split at 100 samples per stratum, yielding 635 samples. Separately, to build the training set, we cap the train split at a larger budget---2k samples per stratum in the base configuration and 4k in our best configuration---with validation capped at 500 per stratum.

\section{Evaluation Protocol}
\label{app:eval}

To ensure reliable benchmarks from all models, we run between 3 and 10 seeds for each model, except for the Claude and GPT models as those APIs either place limitations on or do not allow the seed parameter. We also parameter-tune each model for the best temperature and top-p values prior to evaluation, to ensure we accurately capture best performance.

We use the vLLM reasoning parser to split the reasoning for each model from the structured generation. This ensures that smaller models like \texttt{Qwen3-1.7B} are able to reason and are not constrained from the start by the structured generation limits.

Post-generation, we process the model responses through validation filters, including normalizing register names to lowercase. We find the base \texttt{Qwen3-1.7B} model sometimes duplicates registers in an effort to have multiple options for accuracy; these registers are removed in their entirety.

\begin{table}[htbp]
\centering
\caption{Evaluation metrics and definitions.}
\label{tab:evaluation-metrics}
\begin{tabular}{ll}
\hline
\textbf{Metric} & \textbf{Definition} \\
\hline
Valid Response Rate & LLM produces valid output \\
Register Coverage & Number of expected registers predicted \\
Knowable Register Accuracy & Correct prediction / knowable \\
Unknowable Classification Accuracy & LLM correctly determines whether a register is unknowable \\
Unknowable Value Accuracy & Correct prediction / unknowable \\
\hline
\end{tabular}
\end{table}

\section{Training Details}
\label{app:train}

We explore several experimental variables: the invalid-response penalty (0, $-0.5$, and $-1.0$ on the reward score for a malformed response); the maximum sample size per stratum per task, from a baseline of 2k per stratum ($\sim$12k per task) up to the full dataset; and the weighting of the individual metrics aggregated into the reward. Our experiments are primarily conducted on data subsets ranging from 20k to 100k samples, allowing us to tune training parameters efficiently and iterate quickly.

Nemotron training uses a four-node non-collocated architecture with dedicated vLLM rollout workers and Megatron policy-optimization workers to address node-level memory constraints. Because generation and optimization are performed by separate systems, we apply importance-sampling correction to account for mismatch between the rollout and training policies. Enabling importance sampling and masking sequences with excessive log-probability error ($>1.5$) produces stable early learning.

Later in training, we observe drift caused by adapter-growth-dependent mismatch between vLLM and Megatron. To mitigate this, we increase the reference-policy KL penalty from 0.01 to 0.03 and replace the constant learning rate with cosine decay toward $5\times10^{-7}$. These changes stabilize the mismatch, reduce batch starvation from sequence masking, and allow learning to continue through the full 347-step run.

\section{Reward Function Details}
\label{app:reward}

We expand the reward summarized in Section~\ref{sec:method}. Let $G$ be the set of ground-truth registers for a basic block and $P$ the set of registers the model predicts. For a register $r$, let $m_r = \mathbbm{1}[\hat{v}_r = v_r]$ indicate a correct value prediction, $u^\star_r \in \{0,1\}$ mark whether $r$ is unknowable in the ground truth, and $\hat{u}_r$ be the model's knowable/unknowable label.

Per-register value scores $s_r$ (over all $r \in G$) and classification scores $c_r$ (over predicted registers $r \in G \cap P$) are
\begin{equation}
  s_r =
  \begin{cases}
    -0.5 & r \notin P \quad\text{(missing)}\\
    2m_r - 1 & r \in P,\ u^\star_r = 0 \quad\text{(knowable)}\\
    1.1\,m_r & r \in P,\ u^\star_r = 1 \quad\text{(unknowable)}
  \end{cases}
  \qquad
  c_r = 2\,\mathbbm{1}[\hat{u}_r = u^\star_r] - 1.
\end{equation}
Let $\bar{s}$ and $\bar{c}$ denote the mean per-register value and classification scores:
\begin{equation}
  \bar{s} = \frac{1}{|G|}\sum_{r \in G} s_r,
  \qquad
  \bar{c} = \frac{1}{|G \cap P|}\sum_{r \in G \cap P} c_r.
\end{equation}
Each component passes its mean through a $2^{x}-1$ transform, which bounds the penalty floor at $-0.5$ and rewards higher accuracy superlinearly:
\begin{equation}
  S_{\text{reg}} = 2^{\bar{s}} - 1,
  \qquad
  S_{\text{unk}} = 2^{\bar{c}} - 1,
\end{equation}
and the two combine into the final reward $S_{\text{final}} = 0.7\,S_{\text{reg}} + 0.3\,S_{\text{unk}}$. A response that fails JSON or schema validation instead receives a fixed penalty (default $-0.5$) in place of $S_{\text{final}}$.